\documentstyle[twoside,psfig,amsmath,amsfonts,amssymb,times]{article}

\catcode`\@=11
\long\def\@makefntext#1{
\protect\noindent \hbox to 3.2pt {\hskip-.9pt  
$^{{\eightrm\@thefnmark}}$\hfil}#1\hfill}		

\def\thefootnote{\fnsymbol{footnote}}
\def\@makefnmark{\hbox to 0pt{$^{\@thefnmark}$\hss}}	
	
\def\ps@myheadings{\let\@mkboth\@gobbletwo
\def\@oddhead{\hbox{}
\rightmark\hfil\eightrm\thepage}   
\def\@oddfoot{}\def\@evenhead{\eightrm\thepage\hfil
\leftmark\hbox{}}\def\@evenfoot{}
\def\sectionmark##1{}\def\subsectionmark##1{}}

\oddsidemargin=\evensidemargin
\renewcommand{\thefootnote}{\fnsymbol{footnote}}

\newcounter{sectionc}\newcounter{subsectionc}\newcounter{subsubsectionc}
\renewcommand{\section}[1] {\vspace{12pt}\addtocounter{sectionc}{1} 
\setcounter{subsectionc}{0}\setcounter{subsubsectionc}{0}\noindent 
	{\tenbf\thesectionc. #1}\par\vspace{5pt}}
\renewcommand{\subsection}[1] {\vspace{12pt}\addtocounter{subsectionc}{1} 
	\setcounter{subsubsectionc}{0}\noindent 
	{\bf\thesectionc.\thesubsectionc. {\kern1pt \bfit #1}}\par\vspace{5pt}}
\renewcommand{\subsubsection}[1] {\vspace{12pt}\addtocounter{subsubsectionc}{1}
	\noindent{\tenrm\thesectionc.\thesubsectionc.\thesubsubsectionc.
	{\kern1pt \tenit #1}}\par\vspace{5pt}}
\newcommand{\nonumsection}[1] {\vspace{12pt}\noindent{\tenbf #1}
	\par\vspace{5pt}}

\newcounter{appendixc}
\newcounter{subappendixc}[appendixc]
\newcounter{subsubappendixc}[subappendixc]
\renewcommand{\thesubappendixc}{\Alph{appendixc}.\arabic{subappendixc}}
\renewcommand{\thesubsubappendixc}
	{\Alph{appendixc}.\arabic{subappendixc}.\arabic{subsubappendixc}}

\renewcommand{\appendix}[1] {\vspace{12pt}
        \refstepcounter{appendixc}
        \setcounter{figure}{0}
        \setcounter{table}{0}
        \setcounter{lemma}{0}
        \setcounter{theorem}{0}
        \setcounter{corollary}{0}
        \setcounter{definition}{0}
        \setcounter{equation}{0}
        \renewcommand{\thefigure}{\Alph{appendixc}.\arabic{figure}}
        \renewcommand{\thetable}{\Alph{appendixc}.\arabic{table}}
        \renewcommand{\theappendixc}{\Alph{appendixc}}
        \renewcommand{\thelemma}{\Alph{appendixc}.\arabic{lemma}}
        \renewcommand{\thetheorem}{\Alph{appendixc}.\arabic{theorem}}
        \renewcommand{\thedefinition}{\Alph{appendixc}.\arabic{definition}}
        \renewcommand{\thecorollary}{\Alph{appendixc}.\arabic{corollary}}
        \noindent{\tenbf Appendix \theappendixc #1}\par\vspace{5pt}}
\newcommand{\subappendix}[1] {\vspace{12pt}
        \refstepcounter{subappendixc}
        \noindent{\bf Appendix \thesubappendixc. {\kern1pt \bfit #1}}
	\par\vspace{5pt}}
\newcommand{\subsubappendix}[1] {\vspace{12pt}
        \refstepcounter{subsubappendixc}
        \noindent{\rm Appendix \thesubsubappendixc. {\kern1pt \tenit #1}}
	\par\vspace{5pt}}

\newcommand{\aseff}{{\cal A}_{\textrm{eff}}}
\newcommand{\textlineskip}{\baselineskip=13pt}
\newcommand{\smalllineskip}{\baselineskip=10pt}

\def\eightcirc{
\begin{picture}(0,0)
\put(4.4,1.8){\circle{6.5}}
\end{picture}}
\def\eightcopyright{\eightcirc\kern2.7pt\hbox{\eightrm c}}

\newcommand{\publisher}[2]{{\begin{center}\footnotesize\smalllineskip 
	Received #1\\
	Revised #2
	\end{center}
	}}

\def\abstracts#1#2#3{{
	\centering{\begin{minipage}{4.5in}\footnotesize\baselineskip=10pt
	\parindent=0pt #1\par 
	\parindent=15pt #2\par
	\parindent=15pt #3
	\end{minipage}}\par}} 

\def\keywords#1{{
	\centering{\begin{minipage}{4.5in}\footnotesize\baselineskip=10pt
	{\footnotesize\it Keywords}\/: #1
	\end{minipage}}\par}}

\renewenvironment{thebibliography}[1]
        {\frenchspacing
	 \ninerm\baselineskip=11pt
         \begin{list}{\arabic{enumi}.}
        {\usecounter{enumi}\setlength{\parsep}{0pt}     
	 \setlength{\leftmargin 12.7pt}{\rightmargin 0pt} 
         \setlength{\itemsep}{0pt} \settowidth
	{\labelwidth}{#1.}\sloppy}}{\end{list}}

\newcounter{itemlistc}
\newcounter{romanlistc}
\newcounter{alphlistc}
\newcounter{arabiclistc}

\newcommand{\fcaption}[1]{
        \refstepcounter{figure}
        \setbox\@tempboxa = \hbox{\footnotesize Fig.~\thefigure. #1}
        \ifdim \wd\@tempboxa > 5in
           {\begin{center}
        \parbox{5in}{\footnotesize\smalllineskip Fig.~\thefigure. #1}
            \end{center}}
        \else
             {\begin{center}
             {\footnotesize Fig.~\thefigure. #1}
              \end{center}}
        \fi}

\newcommand{\tcaption}[1]{
        \refstepcounter{table}
        \setbox\@tempboxa = \hbox{\footnotesize Table~\thetable. #1}
        \ifdim \wd\@tempboxa > 5in
           {\begin{center}
        \parbox{5in}{\footnotesize\smalllineskip Table~\thetable. #1}
            \end{center}}
        \else
             {\begin{center}
             {\footnotesize Table~\thetable. #1}
              \end{center}}
        \fi}

\def\@citex[#1]#2{\if@filesw\immediate\write\@auxout
	{\string\citation{#2}}\fi
\def\@citea{}\@cite{\@for\@citeb:=#2\do
	{\@citea\def\@citea{,}\@ifundefined
	{b@\@citeb}{{\bf ?}\@warning
	{Citation `\@citeb' on page \thepage \space undefined}}
	{\csname b@\@citeb\endcsname}}}{#1}}

\newif\if@cghi
\def\cite{\@cghitrue\@ifnextchar [{\@tempswatrue
	\@citex}{\@tempswafalse\@citex[]}}
\def\citelow{\@cghifalse\@ifnextchar [{\@tempswatrue
	\@citex}{\@tempswafalse\@citex[]}}
\def\@cite#1#2{{$\null^{#1}$\if@tempswa\typeout
	{IJCGA warning: optional citation argument 
	ignored: `#2'} \fi}}

\def\pmb#1{\setbox0=\hbox{#1}
	\kern-.025em\copy0\kern-\wd0
	\kern.05em\copy0\kern-\wd0
	\kern-.025em\raise.0433em\box0}

\def\fnt#1#2{\footnotetext{\kern-.3em
	{$^{\mbox{\scriptsize #1}}$}{#2}}}

\def\ps@myheadings{%
    \let\@oddfoot\@empty\let\@evenfoot\@empty
    \def\@evenhead{\slshape\leftmark\hfil}
    \def\@oddhead{\hfil{\slshape\rightmark}}
    \let\@mkboth\@gobbletwo
    \let\sectionmark\@gobble
    \let\subsectionmark\@gobble
    }
\font\tenrm=cmr10
\font\tenit=cmti10 
\font\tenbf=cmbx10
\font\bfit=cmbxti10 at 10pt
\font\ninerm=cmr9

\font\eightrm=cmr8

\def\qed{\hbox{${\vcenter{\vbox{		    
   \hrule height 0.4pt\hbox{\vrule width 0.4pt height 6pt
   \kern5pt\vrule width 0.4pt}\hrule height 0.4pt}}}$}}

\renewcommand{\thefootnote}{\fnsymbol{footnote}}    

\def\bsc{{\sc a\kern-6.4pt\sc a\kern-6.4pt\sc a}}  	
\def\bflatex{\bf L\kern-.30em\raise.3ex\hbox{\bsc}\kern-.14em 
T\kern-.1667em\lower.7ex\hbox{E}\kern-.125em X} 

\begin{document}
\setlength{\textheight}{7.7truein}  

\thispagestyle{empty}

\markboth{\protect{\footnotesize\it Variational Monte Carlo for Microscopic Cluster Models}}{\protect{\footnotesize\it Variational Monte Carlo for Microscopic Cluster Models}}

\normalsize\textlineskip

\setcounter{page}{1}


\vspace*{0.88truein}

\centerline{\bf Variational Monte Carlo for Microscopic Cluster Models}
\vspace*{0.035truein}
\centerline{\footnotesize Theodoros Leontiou}
\baselineskip=12pt
\centerline{\footnotesize\it Department of Physics, UMIST
,P.O Box 88}
\baselineskip=10pt
\centerline{\footnotesize\it Manchester M60 1QD, 
UK }
\centerline{\footnotesize\it E-mail: leontiou@theory.phy.umist.ac.uk}

\vspace*{15pt}          
\centerline{\footnotesize Niels R. Walet}
\baselineskip=12pt
\centerline{\footnotesize\it Department of Physics, UMIST, P.O Box 88}
\baselineskip=10pt
\centerline{\footnotesize\it Manchester M60 1QD, UK}
\centerline{\footnotesize\it E-mail: niels.walet@umist.ac.uk}
\vspace*{0.225truein}
\publisher{(received date)}{(revised date)}

\vspace*{0.25truein}
\abstracts{We examine the application of the Variational Monte Carlo
(VMC) method to a cluster model for halo nuclei. Particular attention
is paid to the error estimate in the presence of correlations in the
underlying random walk. We analyse the required steps for a reliable
application of the VMC in the case of a complicated many-body problem
such as the direct solution of the nuclear hamiltonian with realistic
interactions. We also examine the possibility of variance reduction
through the `zero variance principle', paying particular attention to
the complexity of the many-body problem.}{}{}

\vspace*{5pt}


\vspace*{1pt}\textlineskip	
\section{Introduction and Objectives}		
\vspace*{-0.5pt}
\noindent
One of the great difficulties in nuclear physics, is the numerical
solution of the many-body Schr\"odinger equation with a complicated
state-dependent two-body potential. Although the theoretical
background is fairly simple, a numerical evaluation of the matrix
elements is required, due to the complexity of the many-body
integrals. The variational Monte Carlo (VMC) \cite{kalos,guar:guar1}
is a well known method that can be used to numerically evaluate
expectation values, particularly when the number of variables is
large, such as in the many-body problem.

Our interest lies in the nuclear many-body problem and in particular
in the area of light halo nuclei. Of particular importance is the
presence of correlations which result from the nucleon-nucleon
interaction. For this reason we require a direct solution of the
Schr\"odinger equation. The final wavefunction is constructed by
correlating a starting wavefunction that describes a simple
non-interacting system, i.e., a product wavefunction or Slater
determinant. The full wavefunction is then given by the action of a
correlation operator on this starting function. The exact form of such
a wavefunction is provided by the Coupled Cluster method (CCM)
\cite{kummel,bish:bish1}. In our approximation the correlation
operator is composed of the product of linear and non-linear
parts. The linear part (where the coefficients are determined from the
eigenvalue problem) describes the long-range behaviour of the system
and can be obtained by retaining only the linear terms in the CCM
expansion of the wavefunction. The non-linear part is a Jastrow-type
correlation operator
\cite{jastrow:jastrow1,jastrow:jastrow2} that includes the short-range
correlations appropriate to the short-range behaviour of the nuclear
force. This central operator is parametrised in terms of a number of
parameters that appear non-linearly in the variational principle.

One important issue in light nuclei is translational invariance, which
can drastically effect the expectation value if not properly taken
into account. For this reason the correlation operators depend on the
inter-particle distances only and are thus invariant under
translations. Since the physical picture of simple halo nuclei can be
assumed to be that of an alpha-particle (or a number of
alpha-particles) accompanied by a number of weakly-bound nucleons, we
can use as a starting function the uncorrelated product of the
alpha-particle(s) wavefunction(s) with the valence nucleon(s)
ones. The alpha-particle wavefunction \cite{bish:bish3} is obtained by
correlating a starting function which is the product of harmonic
oscillator wavefunctions. This can be separated into relative and
centre-of-mass parts and can thus preserve translational
invariance. The additional nucleons in the starting function are then
assigned coordinates relative to the alpha-particle(s)
centre(s)-of-mass and the formulation thus preserves translational
invariance. A number of additional non-linear variational parameters
enter the starting function that are related to the average separation
between the individual constituents. Apart from the right quantum
numbers the antisymmetry condition implementing the Pauli exclusion
principle must be imposed. We make use of a scalar correlation
operator that is invariant under any exchange of particle labels and
thus all symmetry requirements reside in the starting function.

Therefore, a variational solution of the many-body problem can result
in a large parameter space, where the expectation value of the
Hamiltonian involves a large number of complicated integrals. The VMC
is an efficient method that can be used for the `statistical'
evaluation of such expectation values. More sophisticated methods that
have been successfully applied in the nuclear many-body problem are
the Diffusion Monte Carlo (DMC)
\cite{dmc} and Green's function Monte Carlo (GMC) methods
\cite{monte2,monte3,monte4,monte6}. These are rather complicated
methods that can provide statistically exact estimates for the
expectation value of a Hamiltonian provided a starting wavefunction
(usually obtained with VMC) is provided. However, the complexity of
the many-body problem restricts for the time being the application of
such methods up to medium-mass nuclei and for simple configurations
and is not always easy to apply to halo nuclei. Although these
methods are accurate they can only provide the numerical value of the
expectation value and the particular structure provided by the
wavefunction is not obvious. On the other hand the VMC provides an
approximation to the expectation value (according to the choice for
the trial wavefunction) and can serve as a starting point for further
investigation. This way the structure required for the best
description of the physical system in question can be easily
identified.

The complexity of the many-body problem requires special attention
rather than a naive application of the VMC method. This is mainly due
to the presence of correlations in the random walk underlying the
method. Apart for complicating the error estimate, such correlations
can impose limitations on the applicability of the VMC method. In this
paper we provide a detailed examination of the VMC method as applied
to a realistic nuclear many-body problem. We investigate the presence
of correlations and the technical issues involved in the evaluation of
a reliable error estimate. We shall make use of a number of
statistical concepts, most of which can easily be found in the
literature such as
\cite{kalos}. In addition we describe a possible variance reduction technique
that potentially can be used to improve the performance of VMC.

\setcounter{footnote}{0}
\renewcommand{\thefootnote}{\alph{footnote}}

\section{The Variational Problem}
\noindent
The basic equation in the VMC is the time-independent Schr\"odinger
equation,
\begin{equation}\label{sch}
H(x)\Psi(x)=E\Psi(x),
\end{equation}
where in general we approximate $\Psi(x)$ as
\begin{equation}\label{wavef}
\Psi(x)=\sum_{n}C_{n}f^{n}(x).
\end{equation}
The wavefunction is expanded in terms of a set of normalizable
trial functions linear in the coefficients $\{C_{n}\}$ and $H$ is
the hamiltonian. In general, $x$ denotes the set of coordinates
appropriate for the many-body hamiltonian. However, for simplicity
spin-isospin degrees of freedom are ignored here. Multiplying
equation (\ref{sch}) on the left by the complex conjugate
wavefunction and integrating over the appropriate variable, the
equation takes the form
\begin{equation}
\sum_{n}C^{*}_{k}\left(\int f^{*}_{k}\hat{H}f_{n}d\Omega\right)
C_{n} =E \sum_{n}C_{k}^{*}\left(\int f^{*}_{k}f_{n}d\Omega\right)
C_{n}
\end{equation}
where $d\Omega$ is the volume element. The above equation can now
be written as
\begin{equation}\label{eqn}
E=\frac{\sum_{n}C^{*}_{k}{\cal H}_{kn}C_{n}}{\sum_{n}C_{k}^{*}{\cal
N}_{kn}C_{n}},
\end{equation}
where ${\cal H}_{kn}$ and ${\cal N}_{kn}$ represent the
Hamiltonian and overlap matrix elements with
\begin{eqnarray}\label{helement}
{\cal H}_{kn}&=&\int f_{k}^{*}Hf_{n}d\Omega,\\
{\cal N}_{kn}&=&\int f_{k}^{*}f_{n}d\Omega. \label{nelement}
\end{eqnarray}

The variational principle states that if $E_{0}$ represents the
exact lowest eigenvalue of the particular Hamiltonian then any
estimate, $E$, for $E_0$ obtained from (\ref{eqn}) will be an
upper bound of $E_{0}$. Using this fact a particular form for
$\Psi(x)$ may be optimised by varying the adjustable parameters to
minimise the expectation value of (\ref{eqn}). The best
approximation to the true lowest eigenvalue is obtained by the
variation of the coefficients $C_{n}$. This leads to a set of
coupled equations of the form
\begin{equation}\label{eigen}
\sum_{n}{\cal H}_{kn}C_{n}-E\sum_{n}{\cal N}_{kn}C_{n}=0,
\end{equation}
that constitute a generalised eigenvalue problem. Once the matrix
elements are known, solution of the eigenvalue problem is
straightforward.

\subsection{Error estimate}
\noindent
The matrix elements entering the eigenvalue problem will be evaluated
numerically (as discussed in the next section), leading to an error in
the estimated eigenvalue. In case where an error estimate for
individual matrix elements exists, the total error for the eigenvalue
problem of (\ref{eigen}) can be obtain from the linear perturbation of
the eigenvalue problem
\begin{equation}
\sum_{n}({\cal H}_{kn}+\delta {\cal H}_{kn}) (C_{n} +\delta
C_{n}) =(E+\delta E)\sum_{n}({\cal N}_{kn}+\delta {\cal
N}_{kn})(C_{n} +\delta C_{n}),
\end{equation}
where $\delta E$ is the unknown error. Multiplying on the right
by the same eigenvector and keeping only first order terms leads
to
\begin{equation}\label{error}
\delta E=\frac{1}{C_{k}{\cal N}_{kn}C_{n}}\left( C_{k}\delta
{\cal H}_{kn}C_{n}-E C_{k}\delta {\cal N}_{kn}C_{n}\right),
\end{equation}
with summation convention implied. If the Hamiltonian and overlap
matrix errors were independent of each other the right hand side
of the above equation could have been taken in quadrature giving a
value for the maximum possible error, $\Delta E$, that is the
required error for the numerical calculation :
\begin{equation}\label{correrror}
\Delta E=\frac{1}{C_{k}{\cal N}_{kn}C_{n}}\sqrt{( C_{k}\Delta
{\cal H}_{kn}C_{n})^{2} + (E C_{k}\Delta {\cal N}_{kn}C_{n})^{2}}.
\end{equation}
However, the above equation can lead to an incorrect error estimate
since in reality the errors in the hamiltonian and overlap matrix
elements are likely to be correlated (depending on the method used for
their evaluation). A way of dealing with this problem is through the
covariance matrix, which can be used to define a set of uncorrelated
(independent) observables whose errors can be added in quadrature. The
elements of the covariance matrix are cov$(B_{z}B_{z^{'}})$, defined
as
\begin{equation}
\textrm{cov}(B_{z}B_{z^{'}})=\langle B_z B_{z^{'}}\rangle-\langle
B_z \rangle\langle B_{z^{'}}\rangle \label{cov},
\end{equation}
with $ B_{z}\in \{\{H_z\},\{N_z\}\}$, while the angular brackets
denote an expectation value. This gives a real symmetric matrix of
the form
\begin{equation}
{\cal \bf C}=\left(\begin{array}{cccccc}
\sigma^{2}(H_{11}) &\ldots&
\textrm{cov}(H_{11}H_{nn})&\textrm{cov}(H_{11}N_{11})&\ldots&
\textrm{cov}(H_{11}N_{nn})\\
\vdots & \ddots &  &   &  &\vdots \\
 \textrm{cov}(H_{nn}H_{11})&\ldots&\sigma^{2}(H_{nn})&\textrm{cov}(H_{nn}N_{11})&
\ldots&\textrm{cov}(O_{nn}N_{nn})\\
\textrm{cov}(N_{11}H_{11})&\ldots&\textrm{cov}(N_{11}H_{nn})&\sigma^{2}(N_{11})&\ldots
&\textrm{cov}(N_{11}N_{nn})\\
\vdots&          &&      &\ddots&\vdots    \\
\textrm{cov}(N_{nn}H_{11})&\ldots&\textrm{cov}(N_{nn}H_{nn})&\textrm{cov}(N_{nn}N_{11})&
\ldots &\sigma^{2}(N_{nn})\\
\end{array}\right)
\end{equation}
with dimensions $(2n^2)\times (2n^2)$, where $n\times n$ is the
dimension of the hamiltonian and overlap matrices. The diagonal
elements correspond to the variance of the hamiltonian and overlap
matrix elements $\sigma^2$ that is discussed later. Diagonalising
the covariance matrix is equivalent to obtaining a new set of
uncorrelated observables that each is a linear combination of the
old ones. This new set of uncorrelated observables, $\{Q_{z}\}$,
can then be defined from the eigenvectors of {\cal \bf C} as
\begin{eqnarray}
B_{z}&=&\sum_{k}R_{zk}Q_{k},\\
\sum_{k}{\cal \bf C}_{zk}
\textbf{R}_{kz}&=&\lambda_{z}\textbf{R}_{zz},\\
{\cal \bf C}&=&\textbf{R}\boldsymbol{\Lambda}\textbf{R}^{T}.
\end{eqnarray}
Here $R_{kz}$ is the $z$th eigenvector of ${\cal \bf C}$, while
$\boldsymbol{\Lambda}\equiv\textrm{diag}(\lambda_1,...,\lambda_{2n^2})$.
The new (independent) observables satisfy the condition
\begin{equation}
\textrm{cov}(Q_{k}Q_{z})=\delta_{kz}\lambda_{z},
\end{equation}
i.e. their covariance vanishes and their variance is equal to the
eigenvalues of ${\cal \bf C}$. The error associated with the new
observables is their standard deviation, which according to the
above equation can be found as
\begin{equation}
(\Delta Q_{k})^{2}=\frac{\lambda_{k}}{N-1},
\end{equation}
with $N$ denoting the number of samples taken for the observable.
This way equation (\ref{correrror}) can be written as
\begin{equation}\label{newerror}
\Delta E=\frac{1}{C_{k}{\cal N}_{kn}C_{n}}
\sqrt{\sum_{k}\left(\sum_{ij}C_{i}(\textbf{R}_{H_{ij}k}
-E\textbf{R}_{N_{ij}k})C_{j}\right)^{2}(\Delta Q_{k})^{2}},
\end{equation}
where the errors corresponding to the independent set of
observables, $\{Q_{k}\}$, can be added in quadrature.

\section{Variational Monte Carlo and Error Estimate}
\noindent
The generalised eigenvalue problem of equation (\ref{eigen})
requires the computation of the Hamiltonian and overlap matrix
elements which in general are complicated integrals and one may
wish to choose Monte Carlo sampling to perform the integration.
The matrix elements of equations (\ref{helement}) and
(\ref{nelement}) can be written as
\begin{eqnarray}\label{Hamiltonian}
{\cal H}_{kn}&=&\int dx~
w(x)\left(\frac{f_{k}(x)H(x)f_{n}(x)}{w(x)}\right),\\ \label{normal}
{\cal N}_{kn}&=&\int dx~ w(x)\left(\frac{f_{k}(x)f_{n}(x)}{w(x)}
\right),
\end{eqnarray}
where $w(x)$ plays the role of the probability density function
provided that is definite positive and shares the same symmetries
as the full wave function. The eigenvalue problem leads to
\begin{eqnarray}\nonumber
\sum_{n}C_{n}\int dx~w(x)H_{kn}(x) & = & E\sum_{n}C_{n}\int
dx~w(x)N_{kn}(x,)\\ \label{eigen2}\sum_{n}\langle H_{kn}\rangle
C_{n}& =& E\sum_{n}\langle N_{kn}\rangle C_{n},
\end{eqnarray}
which can be viewed as the expectation value of two distinct
arbitrary functions of a random variable ($H_{kn}$ and $N_{kn}$)
taken from the same probability distribution.

Because of the parametrization of the wavefunction in terms of
the unknown coefficients $C_{n}$ a single probability
distribution function appropriate for each individual matrix
element cannot be extracted. Although in principle is possible to
sample each matrix element based on its own distribution leading
to $n^2$ distinct Monte Carlo algorithms (or $\frac{n(n+1)}{2}$
for a symmetric matrix), time requirements make such a calculation
impractical. A way around this difficulty is to sample all of the
matrix elements based on a single probability distribution.
Although this approach reduces the number of required
simulations, it can lead to a large variance for individual matrix
elements and special care is required.

\subsection{Statistical sampling}
\noindent
In Monte Carlo methods, $\langle H_{kn}\rangle$ or $\langle
N_{kn}\rangle$ are estimated using a large but finite set of
values for $x$ (a finite set of configurations $\{x_{i}\}$ for the
multi-dimensional case) distributed according to $w$. In order to
avoid referring to individual matrix elements we consider the
expectation value of a general operator $O$ which is denoted by
$\langle O\rangle$.

One well-known method of obtaining the required
distribution is the Metropolis algorithm and this is the one we
shall use. In order to generate values of the variable $x$ having
the required distribution the Metropolis algorithm makes use of a
random walk: An initial point (set of coordinates), $x_{0}$, is
chosen and subsequent points are generated in steps by moving in
a random direction, but each time within a prescribed radius from
each individual coordinate. Not all points, however, are kept but
an `accept or reject' method is used. This means that every point
$x_{l}$ in the random walk belonging to the $l_{\textrm{th}}$ step,
is weighted against the previous point, $x_{l-1}$, and is either
chosen or rejected. The decision process is given by the
Metropolis algorithm which generates a Markov chain. In order to
fulfil the criteria for a Markov chain it is essential for each
step that the decision process governing the evolution of the
random walk should not have any dependence on configurations
belonging to previous steps. This means that the decision of
keeping or rejecting point $x_{l}$ should only depend on the
value of the point $x_{l-1}$ and not on any previous points. An
equivalent statement would be to say that the set of values,
$\{O(x_{i})\}$, chosen to form the average should be uncorrelated
with each other.

The random walk provides as with a statistical average that is
different from the exact expectation value, but can be used as an
approximation. The expectation value $\langle O\rangle$
corresponds to the average of the quantity $O$ over an infinite
ensemble of statistically independent trials. This average can in
principle be obtained exactly if integrals similar to
(\ref{Hamiltonian}) and (\ref{normal}) can be evaluated
analytically. The random walk that is actually performed in
simulations provides an average over a finite sequence of
measurements. This sample average or mean will be denoted by
$\bar{O}$. In the case of $N$ samples taken from a distribution,
the expectation value is approximated as
\begin{eqnarray}\nonumber
\langle O\rangle&\approx&\bar{O},\\ \label{rand}
&=&\frac{1}{N}\sum_{i=1}^{N}O(x_{i}),
\end{eqnarray}
where the $x_{i}$ represents the set of appropriate coordinates
that are distributed according to a probability density w($x$).

\subsection{Error estimate and correlations}
\noindent
Provided we have a statistical average composed of $N$
uncorrelated samples taken from an arbitrary probability
distribution, the central limit theorem states that in the limit
of large $N$, the above average follows a normal distribution
with mean $\langle O \rangle$ and variance
\begin{equation}\label{var0}
\sigma^{2}=\frac{1}{N}\sigma_0^2,
\end{equation}
with
\begin{equation}
\sigma_0^2=\langle O^2\rangle-\langle O\rangle^{2}.
\end{equation}

However, in a Monte Carlo sampling we are limited to a finite sample
average $\bar{O}$
and we would like to know how to estimate the error of such an average
which is the deviation from $\langle O\rangle$ (the limiting case
provided by the central limit theorem). The theoretical value of this
error can be obtained by considering an ensemble of independent random
walks, where we can introduce the notion of the expected value of a
sample average, $\langle
\bar{O}\rangle$, as well as that of the expected value of any
single measurement ($O_i\equiv O(x_i)$) corresponding to a
particular step $i$ of the random walks, denoted by $\langle O_i
\rangle$. The demand that the individual measurements belong  
to a Markov chain and must in principle satisfy
\begin{equation}\label{condition1}
\langle O_i \rangle=\langle O_j \rangle=\langle \bar{O} \rangle
=\langle O\rangle,
\end{equation}
meaning that all samples or measurements in the random walk are
independent of their position relative to the others. The fact
that there is a correlation between individual measurements
corresponds to the case where
\begin{equation}\label{condition2}
\langle O_iO_j\rangle\neq\langle O_i\rangle\langle O_j\rangle
\end{equation}
When the above is taken into consideration the variance of the
mean becomes
\begin{eqnarray}\nonumber
\sigma^2(\bar{O})&=&\langle (\bar{O}-\langle O\rangle)^2\rangle\\ \nonumber
&=&\langle (\bar{O}-\langle \bar{O}\rangle)^2\rangle\\ \nonumber
&=&\langle \bar{O}^2\rangle-\langle \bar{O}\rangle^2,\\ 
&=&\frac{1}{N^2}\sum_{ij}\left(\langle O_i O_j\rangle-\langle
O_i\rangle \langle O_j\rangle\right).
\end{eqnarray}
This describes the deviation of the calculated mean from the
theoretical expectation value.

However, in practise it is impossible to have an algorithm which can
generate samples that are completely uncorrelated. This is indeed the
case of the Metropolis algorithm which is used to generate the Markov
chain. By the nature of the algorithm used the individual
measurements are not statistically independent. Such correlations
between statistical measurements have in general two effects. First,
they reduce the number of independent measurements from the total
number of performed measurements, hence the calculation converges more
slowly.  Second, the estimation for the statistical error will have to
incorporate the effect of such correlations otherwise any error
estimate will be an underestimate. As for any numerical method a
correct error estimate is of crucial importance for the validity of
the results. The rest of this section is devoted to the analysis of
the error estimates for correlated data.  Most of the results are
taken from \cite{error1} and
\cite{error2}.

In case that (\ref{condition2}) does not hold the equation for
the variance of $\bar{O}$ reduces to that of equation
(\ref{var0}). However, when (\ref{condition2}) holds the true
variance for the mean can be written as \cite{error1}
\begin{equation}\label{var1}
\sigma^2(\bar{O})=\frac{1}{N}\left[\sigma_0 + 2\sum_{t=1}^{N-1}
\left(1-\frac{t}{N}\right)\rho_t\right],
\end{equation}
where
\begin{eqnarray}\nonumber
\rho_t &\equiv& \langle O_i O_j\rangle-\langle O_i\rangle \langle
O_j\rangle~~~~t=|i-j|,\\ \label{rhot} &=&\langle O_i
O_{i+t}\rangle-\langle O_i\rangle \langle O_{i+t}\rangle.
\end{eqnarray}
Furthermore, the presence of correlations also influences the
value of the covariance between the means of two different
quantities, $O$ and $O^{'}$, since
\begin{eqnarray}\nonumber
\textrm{cov}(\bar{O}\bar{O}^{'})&=&\langle \bar{O} \bar{O}^{'}\rangle -
\langle \bar{O}\rangle\langle \bar{O}^{'}\rangle\\ \nonumber
&=&\frac{1}{N^2}\sum_{ij}\left(\langle O_i O^{'}_j\rangle-\langle O_i\rangle
\langle O^{'}_j\rangle\right)\\
&=&\frac{1}{N}\left[\gamma_0+2\sum_{t=1}^{N-1}
\left(1-\frac{t}{N}\right)\gamma_t\right]\label{cov1},
\end{eqnarray}
where similarly to equation (\ref{rhot}) we define
\begin{equation}
\gamma_t \equiv \langle O_i O_j^{'}\rangle-\langle O_i\rangle
\langle O_j^{'}\rangle~~~~t=|i-j|~.
\end{equation}
An essential assumption for both $\rho_t$ and $\gamma_t$ is the
invariance under `time translations', meaning that only the
separation between the measurements is of importance and not
their place in the random walk (something also included in
(\ref{condition1})), i.e.
\begin{eqnarray}\nonumber
\langle O_iO_j\rangle=\langle O_jO_i\rangle,\\
\langle O^{'}_iO_j\rangle=\langle O_iO^{'}_j\rangle~.
\end{eqnarray}

As we have seen earlier a reliable error estimate requires two
quantities. Firstly, an estimate of the error for each individual
matrix element, something provided by the variance of the mean
for the particular matrix element. Secondly, an estimate for the
covariance between different matrix elements, so that a set of
uncorrelated observables can be obtained. In theory both of these
objects are provided by $\rho_t$ and $\gamma_t$. However, in a
practical simulation only approximate measurements can be made.

\subsection{Estimation of variance and covariance}
\noindent
An estimate for $\rho_t$ and $\gamma_t$ can be obtained through
the auto- and cross-correlation coefficients. The
auto-correlation coefficient, $C_t$ is defined in the case of $N$
samples as
\begin{equation}\label{autocorr}
C_t(O)=\frac{1}{N-1}\sum_{i=1}^{N-t}(O_i-\bar{O})(O_{i+t}-\bar{O}),
\end{equation}
while the cross-correlation coefficient as
\begin{equation}\label{crosscorr}
{\mathbb C}_t(O,O^{'})=\frac{1}{N-1}\sum_{i=1}^{N-t}(O_i-\bar{O})(O^{'}_{i+t}
-\bar{O^{'}}).
\end{equation}
The variable $t$ will be refereed to as the correlation time.
These two coefficients provide biased estimators for $\rho_t$ and
$\gamma_t$, in the sense that they underestimate the actual
values. This is expressed as
\begin{eqnarray}\label{estimator1}
\langle C_t(O)\rangle&=&\rho_t-\sigma^2(O)+\Delta_t,\\
\langle{\mathbb C}_t(O,O^{'})\rangle&=&\gamma_t-\sigma^2(O)+\Delta_t^{'}
\label{estimator2},
\end{eqnarray}
where the terms $\Delta_t$ and $\Delta_t^{'}$ are given as
\begin{eqnarray}
\Delta_t=2\left(\sigma^2(\bar{O})-\frac{1}{N(N-t)}\sum_{i=1}^{N-t}
\sum_{j=1}^{N}\gamma_{ij}\right),\\
\Delta_t^{'}=2\left(\textrm{cov}(\bar{O}\bar{O}^{'})-\frac{1}{N(N-t)}\sum_{i=1}^{N-t}
\sum_{j=1}^{N}\gamma^{'}_{ij}\right),
\end{eqnarray}
with
\begin{eqnarray}
\gamma_{ij}=\langle O_iO_j\rangle - \langle O_i\rangle\langle
O_j\rangle \nonumber, \\
\gamma_{ij}^{'}=\langle O_i O_j^{'}\rangle - \langle O_i\rangle
\langle O_j^{'}\rangle \nonumber.
\end{eqnarray}

However, in most applications the largest correlation time in
$\rho_t$ and $\gamma_t$ is finite, meaning that equations
(\ref{var1}) and (\ref{cov1}) can be approximated by
\begin{eqnarray}
\sigma^2(\bar{O})&\approx&\frac{1}{N}\left[\sigma_0^2 + 2\sum_{t=1}^{T}
\left(1-\frac{t}{N}\right)\rho_t\right],\\
\textrm{cov}(\bar{O},\bar{O}^{'})&\approx&
\frac{1}{N}\left[\textrm{cov}(O,O^{'}) + 2\sum_{t=1}^{T}
2\left(1-\frac{t}{N}\right)\gamma_t\right].
\end{eqnarray}
The meaning of the above approximation for a random walk is that
the correlation between different samples is of finite range in
the sense that $\langle O_iO_j\rangle-\langle O_i\rangle\langle
O_j\rangle$ and $\langle O_iO'_j\rangle-\langle O_i\rangle\langle
O'_j\rangle$ become zero for large enough correlation time
$t=|i-j|$. The parameter $T$ in the above equations represents a
cutoff parameter and is the maximum correlation time that will be
taken into account. The significance of a finite correlation time
is that the biases $\Delta_t$ and $\Delta_t^{'}$ in equations
(\ref{estimator1}) and (\ref{estimator2}) will become arbitrarily
small for sufficiently large number of samples $n$.

Provided that $\frac{T}{N}$ is sufficiently small, the variance
and covariance can be approximated by
\begin{eqnarray}\nonumber
\sigma^{2}(0)&\approx&\frac{\sigma_0^2+2\sum_{t=1}^T C_t}{N}\\
&=&\left(1+2\frac{\sum C_t}{\sigma_0^2}\right)\frac{\sigma_0^2}{N}
\label{var2}\\
\textrm{cov}(O,O^{'})&\approx&\frac{{\mathbb C}_0+2\sum_{t=1}^T
{\mathbb C}_t}{N}\nonumber\\
&=&\left(1+2\frac{\sum{\mathbb C}_t}{{\mathbb C}_0}\right)
\frac{{\mathbb C}_0^2}{N}.
\label{cov2}
\end{eqnarray}
The above equations provide as with a way of measuring the strength of
correlations in a particular simulation through the `normalised'
correlation coefficients, $C_t/\sigma_0^2$ and
${\mathbb C}_t/{\mathbb C}_0$.  These can be obtained for a
particular simulation as a function of the correlation time $t$.

\subsection{Application and results}
\noindent
In order to examine the accuracy of the variance estimate provided by
equation (\ref{var2}) a simple one dimensional problem was
considered. A hamiltonian of the form
\begin{equation}\label{teller}
H(x)=-\frac{1}{2}\frac{\partial^{2}}{\partial
x^{2}}-\frac{1}{\cosh^{2}x}
\end{equation}
was used, where the exact value of the lowest eigenvalue is known and
is equal to $-\frac{1}{2}$. We can approximate this value through a
variational problem in terms of the eigenvalue equation
(\ref{eigen}). A Gaussian non-orthogonal expansion of the form
\begin{equation}\label{gaussians}
f_{n}=e^{-\mu_{n}x^{2}}.
\end{equation}
was used to approximate the wave function with the $\{\mu_{n}\}$
belonging to a geometric series. Since a numerically exact result
can obtained for this approximation, we can also use Monte Carlo
sampling in order to verify the error estimate. The probability
density function was taken to be the square of the first Gaussian.
Since we can obtain the numerically exact values for ${\cal
H}_{kn}$ and ${\cal N}_{kn}$ we can use this to construct an
unbiased estimator for the variance and thus to obtain an
uncorrelated estimate for the variance of each matrix element.
For example the variance for the Hamiltonian matrix elements is
given as
\begin{eqnarray}
\frac{\sigma^2({\cal H}_{kn})}{N}&=&\langle (\bar{{\cal
H}}_{kn}-E_{kn})^2\rangle\nonumber \\ \label{var3}
&\approx& \sum_{i=1}^{n} (\bar{{\cal H}}_{kn}^i-E_{kn})^2,
\end{eqnarray}
where $E_{kn}$ corresponds to the exact value while the summation
is over a number of distinct random walks with $\bar{{\cal
H}}_{kn}^i$ denoting the distinct average obtained in the
$i_{\textrm{th}}$ walk consisting of $N$ samples. The
approximation symbol becomes an equality in the limit of large
$n$. For our simple one dimensional model the convergence is
relatively fast and the results obtained can be referred to as
statistically exact. Simultaneously we can obtain the value of
the variance through the biased estimator of (\ref{var2}) for any
one of the performed averages.

\begin{figure}[htbp]
\vspace*{13pt}
\centerline{\psfig{file=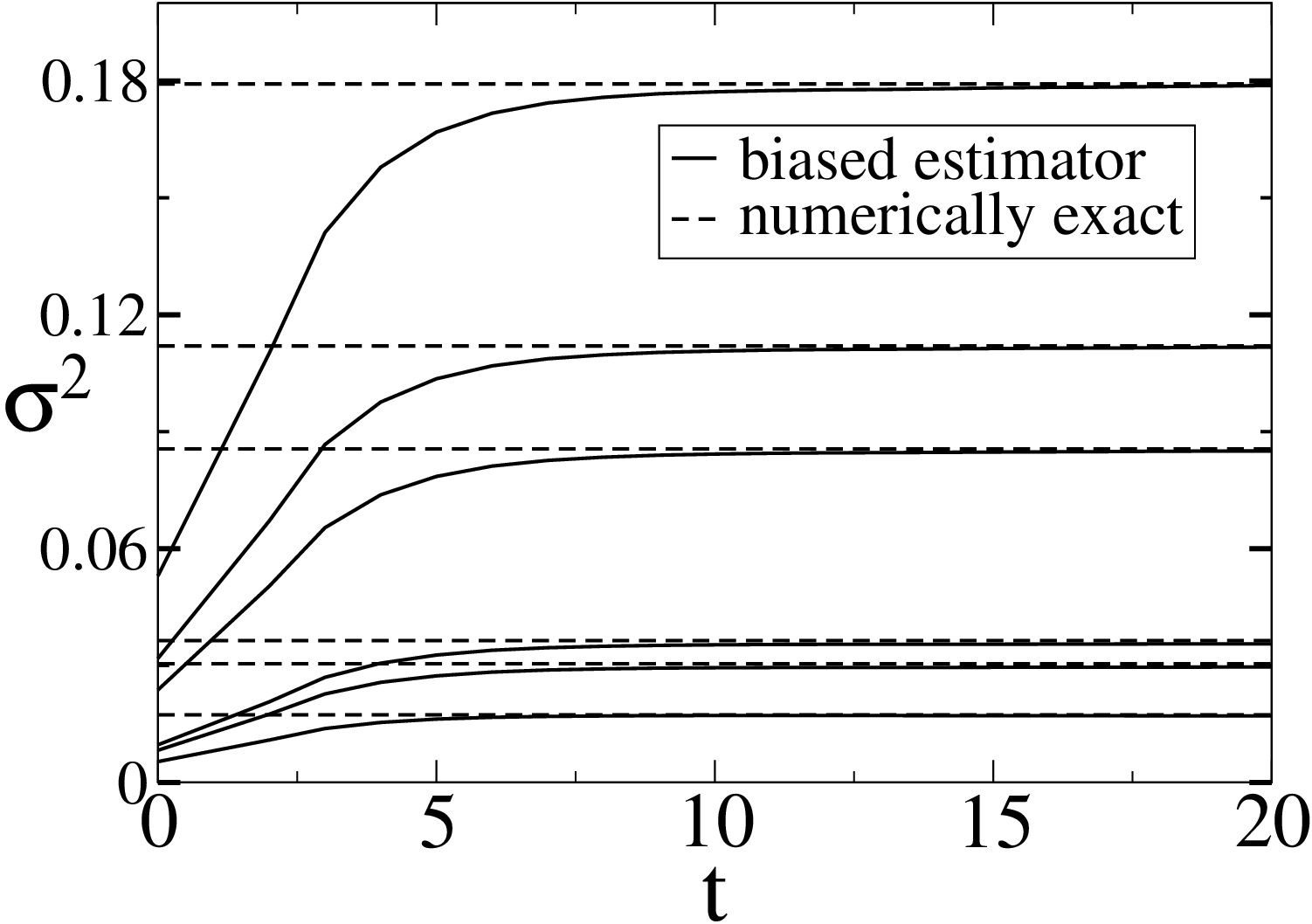,width=10cm,clip=Clip}}
\vspace*{13pt}
\fcaption{\label{varteller} The (biased) variance estimate
for different matrix elements of the Hamiltonian matrix as a
function of the correlation time $t$, for a simple
one-dimensional model. The fact that the biased estimate
approaches a constant value with respect to $t$ indicates that
there is a cutoff in the correlation coefficients (as shown
previously). The dotted lines represent the value for the
variance obtained through an unbiased measurement.}
\end{figure}

The statistically exact value of the variance and that of the biased
estimator depending on the correlation time are shown in figure
\ref{varteller}, which is a demonstration for the accuracy of the
variance estimate. Provided that the correlation time is large enough,
the value for the variance obtained from equation (\ref{var2}) agrees
with that of the statistically exact value of (\ref{var3}). Although
the statistically exact value cannot be obtained directly for
practical calculations (it requires the analytical results), the
biased estimator (\ref{var2}) becomes statistically exact for large
$t$, provided there is a cutoff in the correlation.

When the maximum correlation time is relatively small an unbiased
error estimate can be obtained without having to consider the
correlation coefficient. This is done by rejecting a number of the
samples taken and only considering those which are adequately
separated in simulation time and thus uncorrelated. Since the samples
are consecutively obtained from each other their separation in
simulation time is exactly equivalent to the correlation time. If we
call the maximum correlation time $T$, then each sample is correlated
with samples separated from it by at most $T$ steps. Therefore, if
from the total number of samples obtained only the ones separated by
at least $T$ steps (or random walk moves) are considered, this new set of
samples will be uncorrelated. In the literature these intermediate
moves are usually referred to as `thermalisations'
\cite{kalos}, where is expected that a relatively small number is
adequate. However, for more complicated models, the maximum
correlation time might be too large for either completely
removing the correlations or estimating the effect of the
correlation coefficient without any intermediate moves.

In order to get an idea about the correlation coefficient for the
many-body problem we can consider the linearised approximation of the
many-body wavefunction in terms of correlations operators acting on a
starting function \cite{bish:bish7,bish:bish8}. This provides a
translationally invariant description of the many-body problem and was
applied to a number of closed-shell systems in terms of an
alpha-cluster model \cite{moliner:moliner2}. We firstly consider the
ground-state of the alpha-particle. The many body Hamiltonian is of
the form
\begin{equation}
H=\sum_{i=1}^4\frac{h^2}{2m_i}\nabla^2_i+\sum_{1\leq i<j}^4V(ij),
\end{equation}
where $V(ij)$ is a realistic nucleon-nucleon interaction that is
composed of a sum of Gaussian functions :
\begin{equation}
V(ij)=\sum_{k}c_k\exp(b_k r_{ij}^2),
\end{equation} 
with $r_{ij}$ being the 
distance between the the $i$th and $j$th particles. The wavefunction $\Psi$
is given as
\begin{equation}
\Psi=\hat{F}\Phi_0,
\end{equation}
where $\Phi_0$ represents the wavefunction for the uncorrelated
four-particle system that we take to be the product of harmonic
oscillator wavefunctions. $\hat{F}$ is a correlation operator that is
approximated by a linear expansion of the form
\begin{eqnarray}
\hat{F}=\sum_{i=1}^N \hat{F}_i\Phi_0 =
\left(\sum_{i=1}^N a_i \sum_{i<j}f^i(ij)\right)\prod_{i<j}g(ij),\\
f(ij)=\exp(d_k r_{ij}^2),~~~g(ij)=k_1\exp(-\lambda_1r_{ij}^2)+k_2\exp(-\lambda_2r_{ij}^2).
\end{eqnarray}
This type of wavefunction (refered to as the J-TICI(2) scheme) has been
extensively used for the alpha-particle 
\cite{buendia,niels:niels0,niels:niels1,niels:niels2}
where it was shown to provide an adequate description for the
ground-state properties. The probability density function $w$ was
taken to be the square of one of the components in the expansion of
the wavefunction i.e
\begin{equation}
w=(\hat{F}_0\Phi_0)^2.
\end{equation}
This is a natural choice for the pdf since the wavefunction can always be
written as the product of $w$ and a function $\Psi'$.

\begin{figure}[htbp]
\vspace*{13pt}
\centerline{\psfig{file=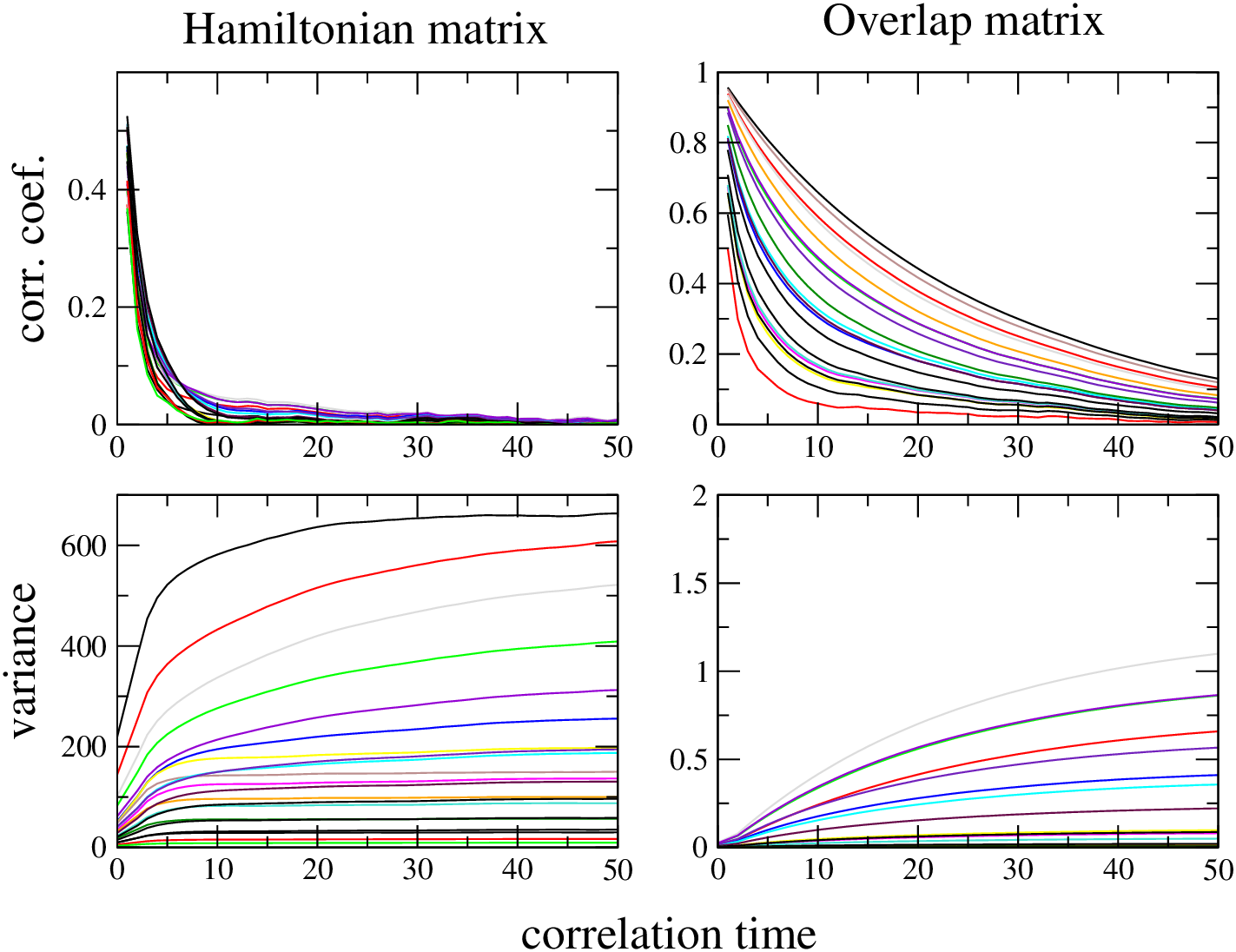,width=12cm,clip=Clip}}
\vspace*{13pt}
\fcaption{\label{corralpha} The variance (lower part) and the 
normalised auto- and cross-correlation coefficients (upper part) as a
function of correlation time. These results are for the matrix
elements of the hamiltonian and overlap matrices of the {\bf
alpha-particle} calculation. The variance graphs start from a 
value belonging to the unbiased estimator and increase as the estimator 
becomes biased by including the effects of the correlation coefficient.}
\end{figure}

The variance of equation (\ref{var2}) and the normalised auto and
cross-correlation coefficients of (\ref{autocorr}) and
(\ref{crosscorr}) were sampled as functions of correlation-time.  This
was done for the matrix elements of both the hamiltonian and overlap
matrices. The result is shown in figure \ref{corralpha}. We can see in
the lower part of the figure that the variance strongly depends on the
correlation coefficient, starting from a minimum (unbiased estimator)
and finally converging. According to the previous analysis this
indicates that despite the fact that the variance depends on the
correlation time, there is a cutoff in the correlation coefficient,
which implies that the dependence on the correlation coefficient will
be over a restricted range of the correlation time (approximately
about the cutoff). This is backed up by sampling the correlation
coefficient (upper graphs of figure
\ref{corralpha}), where we can see that the value of the normalised
correlation coefficient decays rapidly as the correlation time
increases. According to the figure we can safely assume 50 samples as
the value of the cutoff, something that can be compensated by taking
50 intermediate moves in the random walk.

The alpha-particle wavefunction is spherically symmetric and depends on
12 variables (4$\times$3 cartesian coordinates). It is an easy
numerical calculation both in terms of computation time and
complexity. We can go one step further by introducing one or two
additional nucleons, corresponding to the cases of $^5$He and
$^6$He. We shall do so by considering scalar interactions that do not
mix the spatial and spin-isospin terms.  The inclusion of additional
nucleons leads to two complexities. On one hand the wavefunction must
be antisymmetrized, something that does not imply a straight forward
choice for the pdf $w$ and on the other hand the nuclear potential is
composed of several parts including spin, isospin and spin-isospin
exchange terms. The wavefunction has the form
\begin{eqnarray}\nonumber
\Psi&=&\hat{F}{\cal A}\{\Phi_0\chi_{\sigma\tau}\},\\
&=&\sum_ic_i{\cal A}\{\Phi_i\chi_{\sigma\tau}\},
\end{eqnarray}
where the correlation operator has been approximated as before, while
$\chi_{\sigma\tau}$ is the tensor product of the individual spin and
isospin terms. ${\cal A}$ represents an antisymmetrization
operator. The linear correlation operator is invariant under
permutation of particle labels and can be taken outside the
antisymmetrizer. Special attention is required for the
antisymmetrization operator, otherwise the number of terms entering
the wavefunction will be too many for any practical calculation. For
simplicity we do not discuss the quantum numbers corresponding to
angular momentum, total spin and total isospin, having in mind that
the wavefunction must be an eigenstate of good quantum numbers. The
function $\Phi_0$ is not the uncorrelated harmonic oscillator but is
modified to include the motion of the additional nucleon(s) relative
to the alpha-particle. In general the function $\Phi_0$ depends on a
number of non-linear variational parameters that are related to the
average separation between the individual clusters.

\begin{figure}[htbp]
\vspace*{13pt}
\centerline{\psfig{file=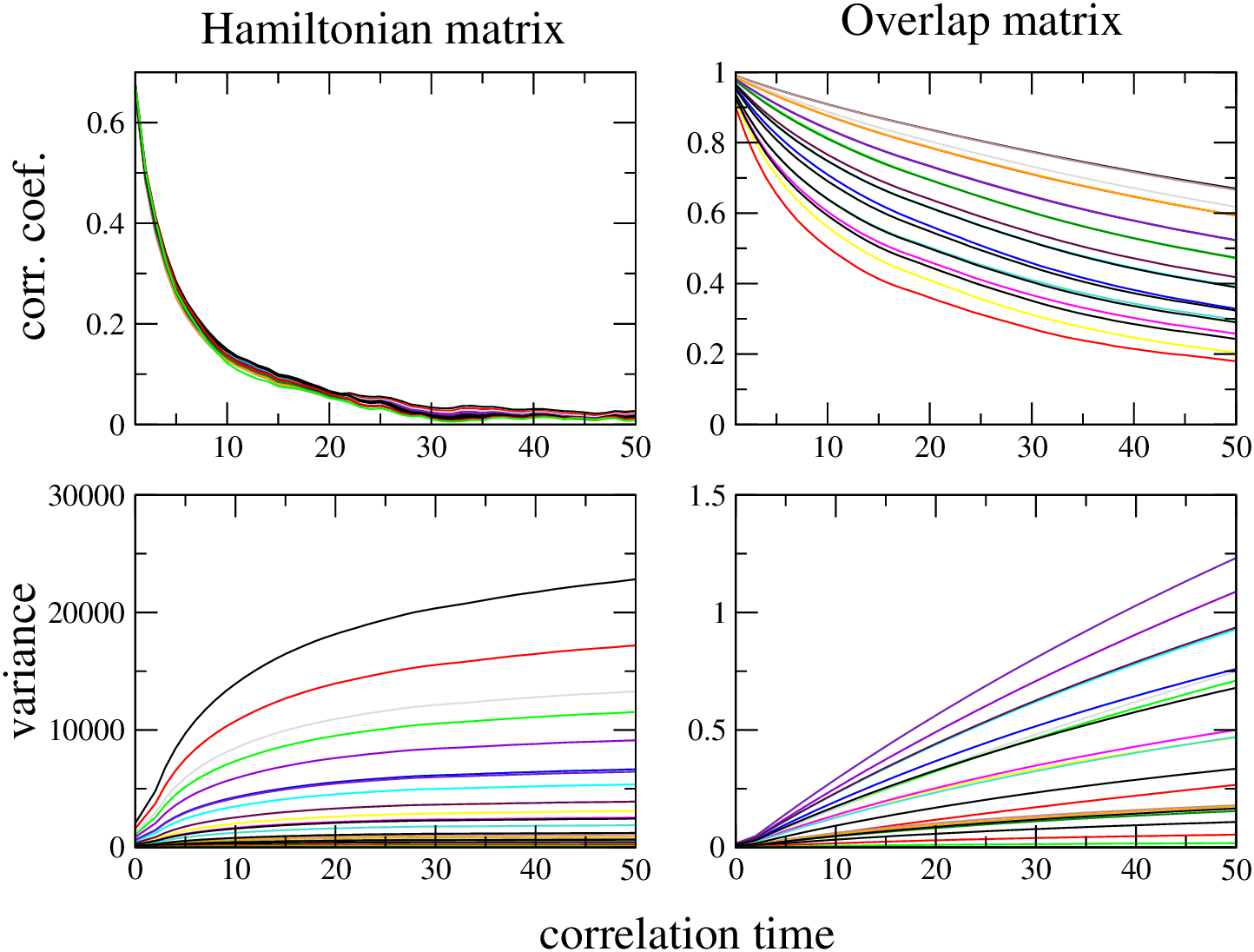,width=12cm}}
\vspace*{13pt}
\fcaption{\label{corr6he} The variance (lower part) and the 
normalised auto- and cross-correlation coefficients (upper part) as a
function of correlation time. These results are for the matrix
elements of the hamiltonian and overlap matrices of the {\bf $^6$He }
calculation. The variance graphs start from a value belonging to the
unbiased estimator and increase as the estimator becomes biased by
including the effects of the correlation coefficient.}
\end{figure}

The norm of the wavefunction can be written as
\begin{eqnarray}\nonumber
\Psi&=&|{\cal A}\Phi|,\\\nonumber
&=&k|\sum_ic_i\aseff\Phi_i|,\\
&=&k\sum_{ij}c_ic_j(\aseff\Phi_i)(\aseff\Phi_j),
\end{eqnarray}
where $k$ is a constant and $\aseff$ is an `effective antisymmetrizer'
that is only composed of a much smaller number of permutations than
${\cal A}$. This provides a relatively simple pdf $w$, for example
\begin{equation}
w=|\aseff\Phi_0|.
\end{equation}
This is not the natural choice as in the case of the alpha-particle
(or any totally symmetric wavefunction) but is rather artificial. We
made use of this type of pdf for going beyond the
alpha-particle. Figure \ref{corr6he} illustrates the variance
estimation for the $^6$He calculation. In the case of the hamiltonian
matrix the correlation coefficient decays rather rapidly and we get a
similar result to the alpha-particle i.e. the estimation of the
variance converges within a relatively small correlation time,
something that can be compensated by taking a number of intermediate
moves. However, in the case of the overlap matrix we see that the
correlation coefficient has a very slow decay so that the variance
estimate does not converge with a reasonable correlation time that can
be compensated with intermediate moves alone. Furthermore, if we go
beyond the $^6$He into heavier systems such as $^8$Be we observe the
same behaviour for the Hamiltonian matrix elements.

\begin{figure}[htbp]
\vspace*{13pt}
\centerline{\psfig{file=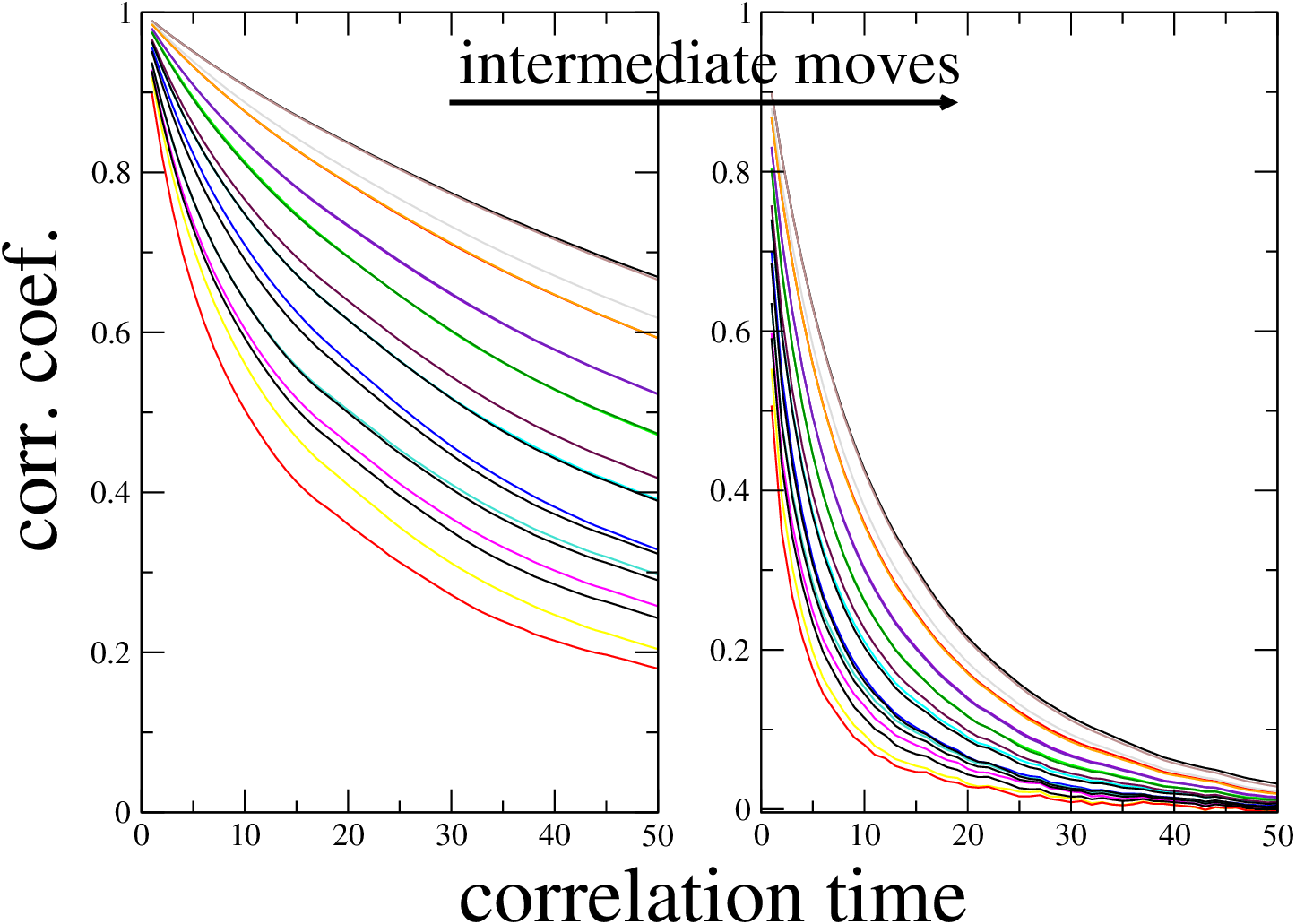,width=8cm}}
\vspace*{13pt}
\fcaption{\label{inter6he} The correlation coefficient before and after
taking intermediate moves (indicated by the arrow). The presence of
intermediate moves introduce a reasonably small cutoff.}
\end{figure}

Therefore, in order to obtain a converging estimate for the variance it
might be more convenient to make use of both intermediate moves and
the unbiased estimator of (\ref{var2}). For example by introducing a
small number of intermediate moves we can force a cutoff to the
correlation coefficient and thus obtain a converging variance through
equation (\ref{var2}). This is illustrated in figure \ref{inter6he}
for the correlation coefficient of the $^6$He overlap matrix: without
any intermediate moves the correlation coefficient decays rather
slowly, while after taking a number of intermediate moves a reasonably
small cutoff is obtained.

It must be noted that the more samples are discarded or the largest
the correlation time is, the more time consuming the simulation
becomes. Therefore, in general the most efficient approach that can
guarantee a correct variance estimate is to make use both of
intermediate moves and the correlation coefficient.

\section{Variance Reduction}

Although correlations can be taken into account or even removed from
the sampling of a particular observable, this can only provide an
algorithm with a reliable error estimate. In order to get an adequate
error a large number of samples are required, the number depending on
the particular observable. The so called `zero-variance principle' is
a way of increasing the efficiency of a Monte Carlo algorithm by
reducing the variance. In general the principle is to replace each
observable to be computed by an improved estimator which has the same
mean but a different variance. The method is described in
\cite{zerovariance} where applications of the zero-variance principle
were shown to be very powerful. This variance reduction technique is
examined in order to establish its usefulness for the case of
many-body cluster models.

\subsection{Method}
\noindent
We consider the expectation value of an observable $O _{kn}$
given as
\begin{equation}
\langle O_{kn}\rangle = \int w(x)O_{kn}(x) dx,
\end{equation}
where as usual $w(x)$ represents a normalised probability density
function (pdf). In our case $\langle O_{kn}\rangle$ corresponds to a particular
matrix element of either the Hamiltonian or overlap matrices,
where for practical purposes all expectation values are over the
same single pdf. This implies that we can optimise $w(x)$ for at
most one of the matrix elements.

We introduce the `renormalized' observable $\tilde{O}_{kn}$ where
\begin{eqnarray}\label{aux}\nonumber
\tilde{O}_{kn}&=&O_{kn}+\frac{\hat{h}\psi_{kn}}{\sqrt{w}},\\
&=&O_{kn}+\bar{O}_{kn} \label{aux2}
\end{eqnarray}
with $\hat{h}$ and $\psi_{kn}$ representing a trial operator and
a trial function appropriate for the expectation of each individual
matrix element. In order for the expectation value of $\tilde{O}_{kn}$
to be the same as that of the original observable the operator
$\hat{h}$ must satisfy the condition
\begin{equation}
\hat{h}\sqrt{w}=0
\end{equation}
This leads to the equation
\begin{equation}\label{fundamental}
\hat{h}\psi_{kn}=(\langle O_{kn}\rangle-O_{kn})\sqrt{w},
\end{equation}
which has the consequence that the error of the calculation,
$\sigma(\tilde{O}_{kn})$, vanishes.
Although in principle equation (\ref{fundamental}) should be satisfied,
in practise an approximate solution can be searched that minimises
the variance of the calculation. The variance for the expectation value
of the new operator of equation (\ref{aux2}) now becomes
\begin{equation}\label{newvar}
\sigma^{2}(\tilde{O}_{kn})=\sigma^{2}(O_{kn})+\sigma^{2}(\bar{O}_{kn})
+2(\langle O_{kn}\bar{O}_{kn}\rangle-\langle O_{kn}\rangle
\langle\bar{O}_{kn}\rangle).
\end{equation}

A possible way to approximate the function $\psi_{kn}$ would be to
consider a finite expansion linear in some coefficients like in
the case of the wavefunction $\Psi$ (equation (\ref{wavef})).
This means that $\psi_{kn}$ can be given as
\begin{equation}\label{auxf}
\psi_{kn}=\sum_{i}A_{i}^{kn}g_{i},
\end{equation} where the
functions $\{g_{i}\}$ are taken to be common for all matrix
elements, while the coefficients $\{A^{kn}_{i}\}$ are determined
by minimising the variance of a particular matrix element, in
this case the observables $\langle \tilde{O}_{kn}\rangle$. The
condition that the variance is minimised can be imposed by
demanding that upon variation of the coefficients $A_{i}^{kn}$
the value of the variance reaches a minimum :
\begin{eqnarray}
\delta \sigma^{2}(\tilde{O}_{kn})&=&0\\
\Rightarrow~~~~~~~~~ \frac{\partial \tilde{O}_{kn}}{\partial
A^{kn}_{i}}&=&0,~~~\forall A^{kn}_{i}.
\end{eqnarray}
This leads for each matrix element to an equation of the form
\begin{equation}\label{zerovar}
\sum_{j}A^{kn}_{j}\Delta_{ij}+K_{i}^{kn}=0,
\end{equation}
where
\begin{eqnarray}\label{delta}
\Delta_{ij}&=&\left\langle\frac{\hat{h}g_{i}\hat{h}g_{j}}{w}\right\rangle
-\left\langle\frac{\hat{h}g_{i}}{\sqrt{w}}\right\rangle
\left\langle\frac{\hat{h}g_{j}}{\sqrt{w}}\right\rangle,\\
\label{kapa} K_{i}^{kn}&=&\left\langle
O_{kn}\frac{\hat{h}g_{i}}{\sqrt{w}}\right\rangle- \left\langle
O_{kn}\right\rangle
\left\langle\frac{\hat{h}g_{i}}{\sqrt{w}}\right\rangle.
\end{eqnarray}
Therefore, for a particular set of functions $\{g_{i}\}$, the coefficients
$\{A^{kn}_{i}\}$ can then be obtained as
\begin{equation}
A^{kn}_{j}=-\sum_{i}(\Delta)^{-1}_{ji}K_{i}^{kn}.
\end{equation}
One difficulty that might arise is when one or more of the eigenvalues
of ${\bf \Delta}$ is near zero (especially when it is zero within the
numerical accuracy of the inversion procedure), in which case solving equations (\ref{zerovar})
by inverting ${\bf \Delta}$ will lead to numerical
instabilities. Diagonalising ${\bf \Delta}$ is equivalent to obtaining
linear superpositions of the functions $\{g_i\}$ that are uncorrelated
with respect to the operator $\frac{h}{\sqrt{w}}$. However, this might
not be possible for the entire set $\{g_i\}$. This problem can be
bypassed by solving (\ref{zerovar}) with singular value decomposition,
which is equivalent to considering a reduced set of functions
$\{g_i\}$.

By substituting the coefficients back to equation (\ref{newvar})
the variance of each particular matrix element becomes
\begin{eqnarray}\label{newvar2}\nonumber
\sigma^{2}(\tilde{O}_{kn})&=&\sigma^{2}(O_{kn})+\sum_{ij}K_{i}^{kn}
A_{i}^{kn},\\
&=& \sigma^{2}(O_{kn})-\sum_{ij}A^{kn}_{i}\Delta_{ij}A^{kn}_{j},
\end{eqnarray}
which is always bound to be less than $\sigma^{2}(O_{kn})$, whatever
the choice for the $g_{i}$'s due to the fact that the eigenvalues of
${\bf \Delta}$ are always positive.

An efficient method of obtaining an appropriate set of functions
$g_{i}$ is to run a number of short Monte Carlo algorithms until a set
that reduces the variance of the calculation efficiently is found. The
set of coefficients $\{A_{i}\}$ can be obtained from a short sampling
and used as input for a larger computation.\\
\newline
{\bf Note}: The error for the zero variance calculation can also be
obtained from a set of uncorrelated errors giving an equation similar to
(\ref{newerror}). The only difference is in the covariance matrix to be
calculated. The new matrix elements are
\begin{eqnarray}
C_{ij}&=&\textrm{cov}(\tilde{B}_{i}\tilde{B}_{j}), \nonumber \\
&=& \textrm{cov}([B_{i}+\bar{B}_{i}][B_{j}+\bar{B}_{j}]), \nonumber \\
&=& \textrm{cov}(B_{i}B_{j})+\sum_{n}A_{n}^{j}K_{n}^{i}+\sum_{n}A_{n}^{i}K_{n}^{j}
+\sum_{nm}A^{i}_{n}A^{j}_{m}\Delta_{nm},
\end{eqnarray}
where $i$ is a collective index representing a particular matrix
element of either the Hamiltonian or overlap matrix
($\{O_{nm},N_{nm}\}\equiv\{B_{i}\}$), while ${\bf \Delta}$ and ${\bf
K}$ are defined in equations (\ref{delta}) and (\ref{kapa}).

\subsection{Application and results}
\begin{figure}[htbp]
\vspace*{13pt}
\centerline{\psfig{file=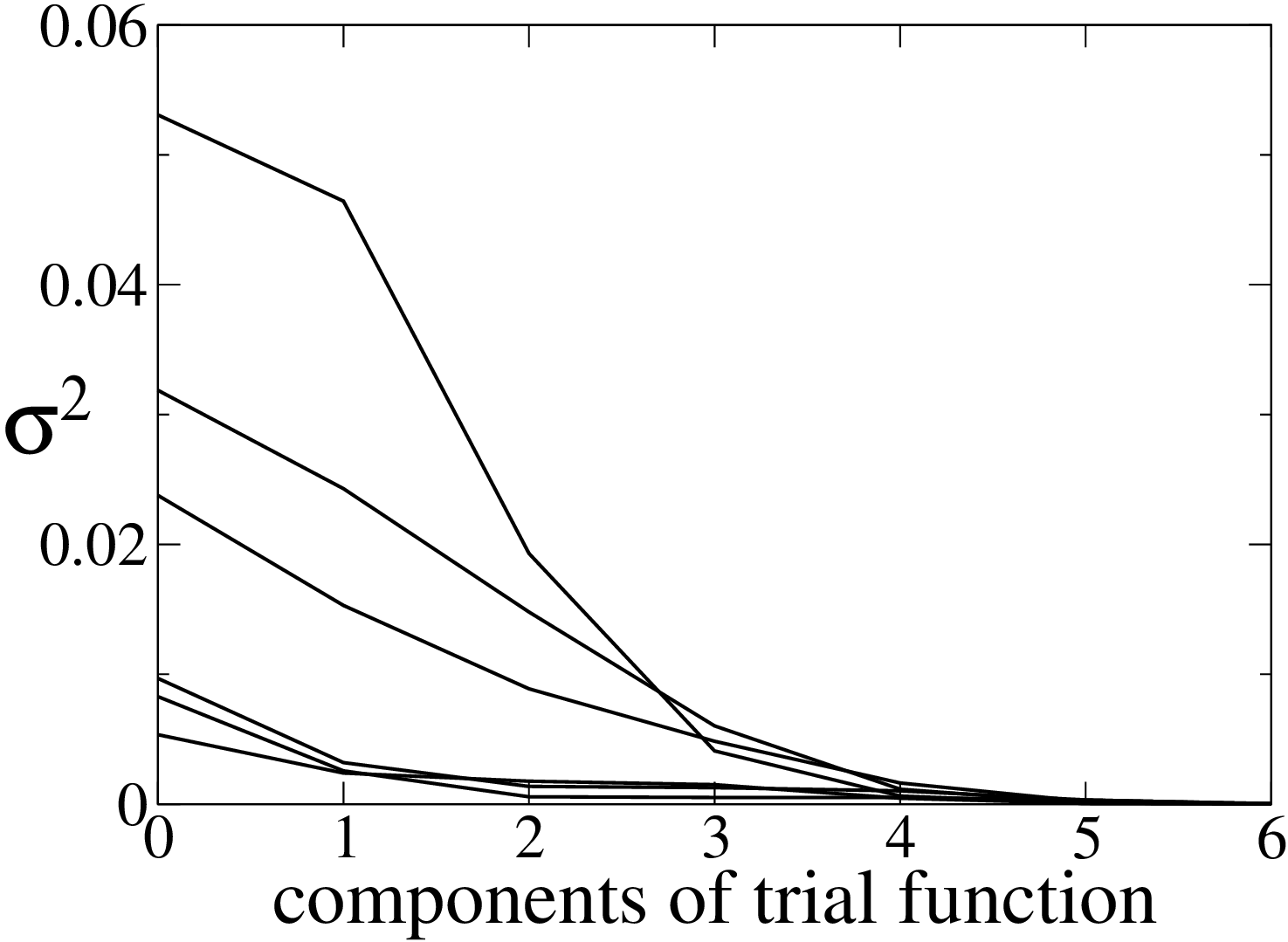,width=10cm}}
\vspace*{13pt}
\fcaption{\label{varred1} The variance of some matrix
elements of the hamiltonian matrix for the one-dimensional problem as
a result of applying the zero variance principle. The variance was
plotted against the number of components used to approximate the trial
function.}
\end{figure}

\noindent
Although the variance reduction technique can in principle reduce the
variance of an observable, for practical calculations we are confined
to the linear approximation of equation (\ref{auxf}). For a
complicated many-body problem the choice of the functions $\{g_{i}\}$
entering (\ref{auxf}) is not at all obvious. For our calculations we
shall use the same set as the one used for the linear eigenvalue
problem. This is the simplest available choice. We can examine the
applicability of this approach through the same one dimensional
problem as before before going to the many-body case. In this case the
integrals of equation (\ref{newvar2}) can be solved in a numerically
exact manner without having to make use of Monte Carlo sampling. The
numerically exact results for the variance of the one-body problem are
shown in figure (\ref{varred1}). In this case the variance reduction
greatly reduces the variance of the matrix elements and can lead to a
zero variance (within the numerical accuracy of the machine used).

\begin{figure}[htbp]
\vspace*{13pt}
\centerline{\psfig{file=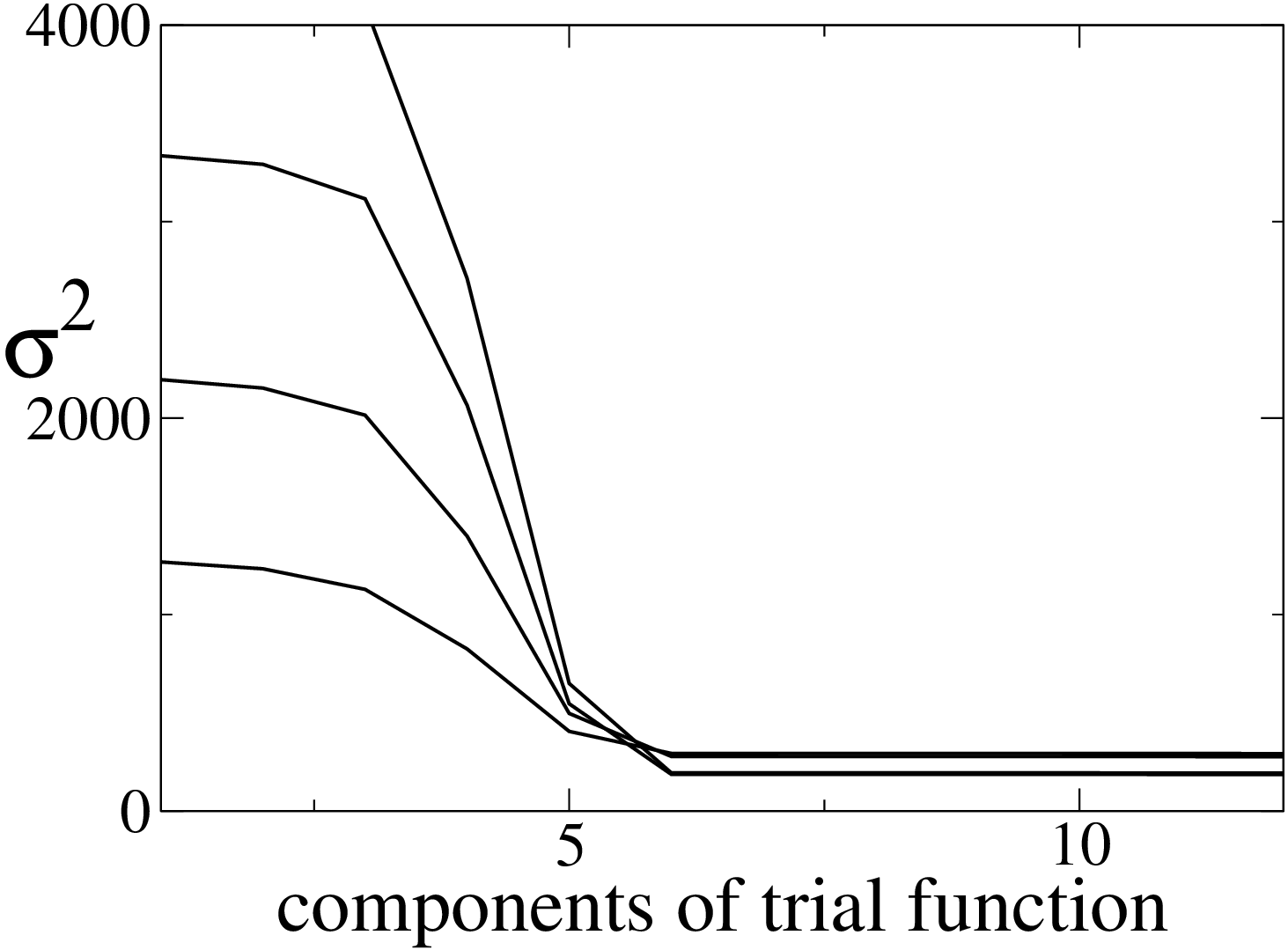,width=10cm}}
\vspace*{13pt}
\fcaption{\label{4hevar} The variance of some matrix
elements of the hamiltonian matrix for the alpha-particle in the
J-TICI(2) approximation as a result of applying the `zero-variance'
principle. The variance is plotted against the number of components
used to approximate the trial function.}
\end{figure}

We can then proceed and apply the same method to the case of the
alpha-particle in the J-TICI(2) approximation, this time using
Monte Carlo sampling. Figure \ref{4hevar} is the variance of a
few of the hamiltonian matrix elements as functions of the number
of linear components used in equation (\ref{auxf}). Although
there is a substantial reduction in the variance, the effect is
not as strong as in the one-dimensional case (not a zero-variance
principle any more)

Although there is a substantial reduction in the variance of the
alpha-particle calculation (about 90$\%$), this is not a reduction
that can be of practical help. Having in mind that the error is given
by the standard deviation (that is the square root of the variance) we
have that its value changes with the number of samples as
$\frac{1}{\sqrt{N}}$. A $90\%$ reduction is (approximately)
equivalent to an error dependence of the form
$\frac{1}{3\sqrt{N}}$. This is not much of an error reduction,
particularly when compare with the simple one-dimensional case.  For
the `zero variance' principle to be valuable we require a radical
variance reduction. Therefore, the variance reduction for the alpha
particle is not sufficient for aiding the numerical calculation. A
similar situation was also observed in the case of the more
complicated systems of $^5$He and $^6$He. Although a variance
reduction was possible it was not of a substantial contribution to the
simulation. Furthermore, the amount of reduction was considerably
lower than that for the alpha-particle. This is not unexpected since
the wavefunction is more complicated and the issue of
antisymmetrization is also present. 

It seems that the complicated structure of the many-body wavefunction
requires a different kind of approximation of (\ref{fundamental}) than
the simple linear one, for any effective reduction of the variance. 

\section{Summary and Conclusions}
\noindent
The variational Monte Carlo method is an important ingredient for
cluster-like models since it allow as to investigate different
structures in the cluster model without having to worry about the
analytical solution of many-body integrals. This requires the
numerical method to be both accurate and precise in the error
estimate. Furthermore, is a starting point for more sophisticated
methods such as GMC.

In our cluster model the VMC in is applied to the generalised
eigenvalue problem, where the total error requires to decorrelate the
individual matrix elements of the hamiltonian and overlap
matrices. This requires knowledge of the covariance matrix, where the
diagonal elements coincide with the variances. This is a straight
forward task that does not impose any difficulty for the many-body
calculations.

We have illustrated that the knowledge of the correlation coefficient
is crucial for a correct estimate of the variance (or covariance),
particularly for complicated systems i.e. the simulation variance is
different from the analytical one due to the presence of correlations.
This requires to make use of a biased variance estimator, where the
bias becomes arbitrarily small as the number of samples
increases. For such an estimator to be practical we require that the
correlation coefficient has a reasonable cutoff, i.e. that a
particular sample in the random walk is only correlated up to a finite
number of previous samples. In the case where the cutoff is relatively
large it can be reduced by discarding a number of samples between the
values taken. The most efficient approach that can guarantee a correct
variance estimate is to make use both of intermediate moves and the
correlation coefficient.

We also examined a variance reduction technique that can be used to
improve the efficiency of a calculation, the so called `zero variance
principle'. This was taken from \cite{zerovariance}. The principle is
in general complicated to apply and we only considered a linear
approximation. This sufficiently decreased the variance of the alpha
particle calculation. For more complicated systems a different kind of
approach is required. In principle we could have looked for a more
complicate approximation than the one at hand, but this is beyond our
purpose. It seems that variance minimisation for the many-body problem
is not a straight forward matter, particularly when antisymmetrization
is involved.

In general we have analysed the application of the VMC method for the
nuclear many-body problem, and in particular for the case of the
generalised eigenvalue problem. We discussed and analysed the required
steps for a reliable error estimate.

\nonumsection{Acknowledgements}
\noindent

The financial support provided by the UK Overseas Research Students
Awards Scheme, a UMIST Graduate Scholarship scheme and the EPSRC for a
project studentship is gratefully acknowledged.

\nonumsection{References}

\end{document}